\documentclass[twocolumn,showpacs,preprintnumbers,amsmath,amssymb]{revtex4}
\usepackage{graphicx}
\begin{document}
\title{Direct observation of large temperature fluctuations during DNA thermal denaturation}
\author{K. S. Nagapriya\footnote[1]{Electronic mail: ksnaga@physics.iisc.ernet.in}, A. K. Raychaudhuri\footnote[2]{Electronic mail: arup@physics.iisc.ernet.in, currently at S.N.Bose National Centre for Basic Sciences, Salt Lake, Kolkata 700098}}        
\address{Department of Physics, Indian Institute of Science,  Bangalore 560 012,  India.}
\author{ Dipankar Chatterji}
\address{Molecular Biophysics Unit, Indian Institute of Science, Bangalore 560 012, India}
\date{\today}
\begin{abstract}
%PACS: 75.30.Vn;68.55.-a
In this paper we report direct measurement of large low frequency temperature fluctuations in double stranded (ds) DNA when it undergoes thermal denaturation transition. The fluctuation, which occurs only in the temperature range where the denaturation occurs,  is several orders more than the expected equilibrium fluctuation. It is absent in single stranded (ss) DNA of the same sequence. The fluctuation at a given temperature also depends on the wait time  and vanishes in a scale of few hours. It is suggested that the large fluctuation occurs due to coexisting denaturated and closed base pairs that are in dynamic equilibrium due to  transition through a potential barrier in the scale of $25-30k_{B}T_{0}$($T_{0}=300K$).
\end{abstract}
\maketitle
\section{Introduction}
The physics of thermal denaturation has been a topic of investigation for the past four decades~\cite{PB1,PB2,Bloomfield,Babri,Garel}. In thermal denaturation, which we call melting transition (also known as helix-coils transition) the hydrogen bonds between the base pairs of the double stranded DNA (dsDNA) break and the dsDNA separates into two single stranded DNA (ssDNA). It occurs over a range of temperature and there are distinct steps which depend on the sequence as well as the length of the DNA (number of base pairs)~\cite{length}. The AT segments with lower binding energy melt at lower temperatures compared to the GC segments. The melting of DNA starts with creation of denaturation bubbles. As the denaturation process proceeds the number of bound pairs decrease. 

In this paper we address the issue of temperature fluctuations during the process of thermal denaturation which we believe has not been addressed before. We present the first direct measurement of the temperature fluctuations  in a number of  double stranded DNA undergoing denaturation. These include heterogeneous polymers  like a 40 kbps T7 bacteriophage, 7250 bps plasmid (both ssDNA and dsDNA), 100bp heteropolymeric dsDNA fragment and  a 100bps homopolymer (...ATAT...). Our experimental results show that the process of thermal denaturation is associated with  large low frequency  fluctuations. These grow as the temperature enters the melting regime and die down once the melting process is complete. At a given temperature in the melting regime, the increased fluctuation shows a time dependence. It first grows and then  decays. 
\begin{figure}
\includegraphics[width=8cm]{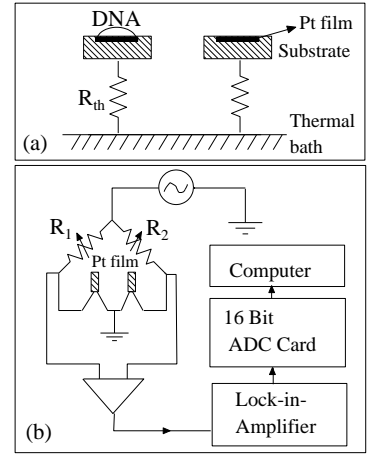}
\caption{\label{fig:figure1} (a) Schematic of the calorimeter, (b) schematic of the noise measurement setup.}
\end{figure}
The basic process of thermal detection is explained in the schematic of figure~1. The calorimeter consists of a substrate (heat capacity $C_{sub}\approx 30mJ/K$) and a thin film of Pt as a thermometer. The calorimeter is connected to a temperature controlled copper thermal bath  (heat capacity $C_{bath}\approx 60J/K$)  through a link of thermal resistance $R_{th}\approx 800 K/W$.The total heat capacity of the system is the heat capacity of the sample $C_s$  plus that of the substrate $C_{sub}$.  But $C_s$ being much less ($\sim 10^{-6} J/K$), the total heat capacity is $\approx C_{sub}$. The calorimeter has a relaxation time $\tau = R_{th}C_{sub}$.  We use  two calorimeters  connected in  a bridge arrangement which enhances the sensitivity of the experiment and also reduces the effect of small thermal drifts. 
\begin{figure}
\includegraphics[width=8cm]{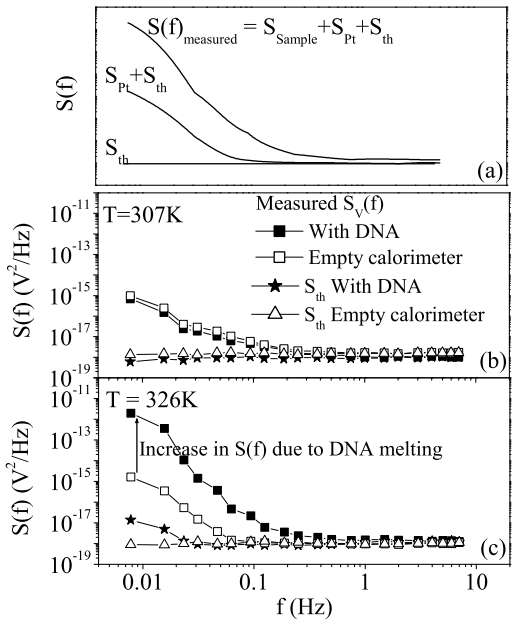}
\caption{\label{fig:figure2} (a) Cartoon of the three contributions to the observed $S_V(f)$. (b) The observed $S_V(f)$ and $S_{th}$ at $T=307K $($<<$ temperature where the denaturation sets in) for empty calorimeter (open symbols) and calorimeter with 7250bp dsDNA (closed symbols) (c) the observed $S_V(f)$ and $S_{th}$ at $T=326K $ for empty calorimeter (open symbols) and calorimeter with 7250bp dsDNA (closed symbols).}
\end{figure}
The sample (DNA) is mounted on the substrate. When the strands separate out the DNA absorbs energy from the calorimeter leading to lowering of the calorimeter temperature in a time scale that is faster than the thermal relaxation time ($\tau$) of the calorimeter. At a longer timescale the calorimeter equilibrates with the heat bath. In the timescale $<\tau$, if the energy exchanged by the DNA with the calorimeter/substrate is $\Delta U$, the temperature change $\Delta T$ of the calorimeter is $\Delta T \approx \Delta U/C_{sub}$. As the denaturation transition proceeds, the process of energy exchange thus leads to a fluctuation in the temperature of the calorimeter (about its average temperature). The  temperature change $\Delta T$ of the calorimeter arising from the energy exchange leads to a  resistance change $\Delta R$ of the thermometer. The thermometer being current biased, this is detected as a voltage change ($\Delta V$). The  voltage fluctuation thus arising from the temperature fluctuation is given by:
\begin{equation}
\label{resfluc2}
\frac{\langle(\Delta V)^2\rangle}{V^2} = (\beta T)^{2}\frac{\langle(\Delta T)^2\rangle}{T^2}
\end{equation}
\noindent
where $V$ is the voltage across the Pt thermometer. $\beta (= \frac{1}{R}\frac{dR}{dT}$)  of the platinum thermometer was determined from the calibration of the thermometers and $\approx 3.8\times 10^{-3}/K$ around 300K. The thermal link of the calorimeter determines the bandwidth of the detection system which is like a low pass filter with a roll off at $f_{c} = \frac {1}{2\pi \tau}$.  

The temperature fluctuations of the calorimeter  can be characterized by the power spectral density (PSD) and the mean square temperature fluctuation  $\langle(\Delta T)^2\rangle$. 
\begin{figure}
\includegraphics[width=8cm]{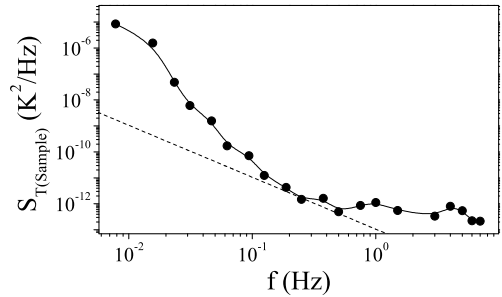}
\caption{\label{fig:figure3}An example of the PSD of the temperature fluctuation $S_{T(Sample)}$ for 7250 bps dsDNA at T=326K (shown by cirles). The PSD of temperature fluctuation due to the temperature fluctuation of the bath is shown by a dotted line.}
\end{figure}
The observed temperature fluctuation though finite is small and a digital signal processing technique ~\cite{Noisetec} was used to eliminate the different background contributions (described later on). Briefly, the voltage fluctuation (and from it the temperature fluctuation) was determined by digitizing the $\Delta V$ of the thermometer as a function of time at a digitization rate of 1024 /sec. To reduce the contribution of the noise of the detector electronics, the voltage fluctuation was measured using a frequency modulation technique where the fluctuation appears as a side band around a carrier frequency($\sim$ 230Hz) ~\cite{Noisetec}. The data is digitally filtered and decimated and the PSD of voltage fluctuation $S_V(f)$ was obtained by a fast fourier transform (FFT). As described later, from the observed $S_V(f)$, the Nyquist noise and the $1/f$ noise of the thermometer (predetermined experimentally using the empty calorimeter) were subtracted out to obtain the PSD of fluctuation from the sample $S_{Sample}(f)$. The mean square fluctuation $\langle(\Delta V )^2\rangle/V^2$ $=$ $\int_{f1}^{f2} S_{Sample}(f)df$ where $f_1$ and $f_2$ are the lower ($\approx 10^{-3}$Hz) and upper ($\approx$ 10Hz) frequency limits. The temperature fluctuation $\langle(\Delta T )^2\rangle/T^2$ was then obtained from $\langle(\Delta V )^2\rangle/V^2$ by equation ~\ref{resfluc2}. The determination of $\langle(\Delta T )^2\rangle/T^2$ through PSD was necessary for proper subtraction of the contribution of Nyquist noise and the thermometer $1/f$ noise.

   A few $\mu$g to a few tens of $\mu$g of a solution of the DNA in water are put on the substrate  using a micropipette and the solution is allowed to evaporate off.  DNA with solvent evaporated off harbours about $66 \%$ bound water~\cite{water} and the conformation of the DNA does not undergo any change. We have observed similar fluctuations near the denaturation temperature when the experiment is carried with DNA in a buffer. We have used DNA with solvent evaporated off because DNA in solution reduces the sensitivity of the experiment due to the thermal mass of the solvent. 

Figure~2 shows the observed  $S_V(f)$ across the Pt film  taken with an empty calorimeter as well as that taken with a  sample of 7250bp dsDNA. It can be seen from figure 2(b)  that at  $T=307K$  both the data are nearly the same. However at $T=326K$ (figure 2(c)) the observed fluctuation with DNA is much larger. This is an example of a typical data. The observed $S_{V}(f)$ can be written (shown in figure~2(a)) as: 
\begin{equation}
S_{V}(f)=S_{th}+S_{pt}+S_{sample}
\end{equation}
Here $S_{th}$ is the white Nyquist noise given by $4k_BTR$ ($\approx 10^{-18} V^2/Hz$) and $S_{pt}$ is predominantly the $1/f$ noise of the Pt film resistor.   The sum $S_{th} + S_{pt}$ (that constitutes the background) is measured by doing the runs on an empty calorimeter and is subtrated out from the measured $S_{V}(f)$ to obtain $S_{sample}$. (Note:  $S_{th}$ can be independently obtained by digitizing the out-of- phase signal from the lock-in amplifier ~\cite{Noisetec}.)

In figure~3 we show an example of the PSD of the temperature fluctuation $S_{T(Sample)}$ (obtained from  $S_{sample}$ using the relation $S_{T(sample)}(f) = (\beta V)^{-2}S_{sample}(f)$)  for the 7250 bp dsDNA at T=326K.  This figure shows that  there is a large temperature fluctuation that has a predominant low frequency component.  In the same figure we also show the expected contribution from the bath fluctuation by a dotted line ~\cite{response}. It can be seen that this is several orders smaller than the fluctuations due to the sample.

The data were taken using two different procedures. In the first procedure, the temperature was stabilized at a particular value and two data sets were taken. Immediately the temperature was raised to the next desired value. At temperatures below that of the onset of melting, the observed voltage fluctuation is mostly the 1/f noise of the thermometer. By slowly increasing the temperature we find the onset temperature where there are additional contributions to the fluctuations. Using this procedure we obtained data  as a function of temperature for fixed wait time $t_w$ at every temperature. In the second method, the thermal bath was stabilized at a temperature where  one sees significant fluctuations (which is known from data got using the first procedure). Data were taken at that temperature at different time intervals  until the additional fluctuations fell below the detection limit. This way, using different wait time $t_w$ we can follow the approach to equilibrium. 

\begin{figure}
\includegraphics[width=8cm]{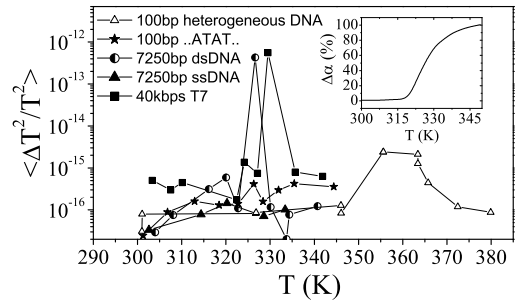}
\caption{\label{fig:figure4} The relative mean  temperature fluctuation $\frac{\langle(\Delta T)^2\rangle}{T^2}$  as a function of temperature in the different DNA samples.}
\end{figure}
\begin{figure}
\includegraphics[width=8cm]{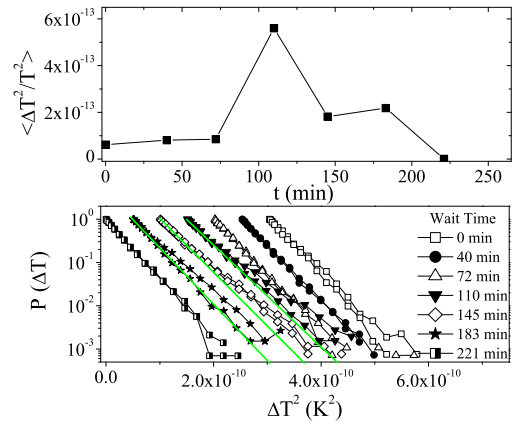}
\caption{\label{fig:figure5} (a) $\frac{\langle(\Delta T)^2\rangle}{T^2}$ as a function of the wait time $t_w$ at 329K for the T7 DNA. (b) the PDF of temperature fluctuation as a function of the wait time $t_w$ at 329K for the T7 DNA. The data are offset for clarity. The deviation from the Gaussian behavior is seen as deviation from the straight lines.}
\end{figure}

In figure~4  we show the $\langle(\Delta T)^2\rangle/T^2$  as a function of temperature for the different DNA samples.  The figure shows that in a certain region in temperature the dsDNA shows a large  fluctuation which is absent in the ssDNA with the same sequence. We find that $\langle(\Delta T)^2\rangle/T^2$ can be rather large, reaching close to 1ppm in the temperature range of interest . In the inset of figure~4  we show the UV absorption data for the T7 DNA. We can see that the melting temperature matches very well with that of increased fluctuation. Thus we can relate the observed temperature fluctuation in the  dsDNA  to the process of thermal denaturation.  The fluctuation is  absent both in the fully zipped and fully unzipped state.  The range in temperature where the excess noise persists depends on the length and heterogeneity, being maximum for the  T7 DNA ($\delta T\approx$ 15K ) and smallest for the 100bps homopolymer ($\delta T \approx$ 5K). 
It can also be seen that while in the 100bp heteropolymer the fluctuation jumps by more than an order of magnitude during melting, in the 100bp homopolymer the change is much lesser. In the case of the T7 DNA, the fluctuation during melting is about three orders higher than that of the background fluctuation. It is clear from these observations that both length and heterogeniety play important roles in determining the magnitude of fluctuation. To summarize, the fluctuation is larger in longer DNA samples  and in heterogenous samples. 

Figure~5(a) shows the $\frac{\langle(\Delta T)^2\rangle}{T^2}$ as a function of the wait time $t_w$ at 329K for the T7 DNA. We define wait time as the time elapsed from when the temperature reached a stable value. This is a typical data set at the onset of the fluctuation. Initially the fluctuation is very small. It increases with time, reaches a maximum and then decreases again. The $t_w$  dependence of the  fluctuation is also clearly seen in the probability distribution function (PDF) of the temperature jumps ($\Delta T$) shown in figure~5(b). The PDF is shown as relative probability vs $(\Delta T)^2$. The data are offset for clarity. For Gaussian distribution these plots should be straight lines.  We see that at $t_w = 0$ the plot is a straight line. With larger $t_w$ there is deviation from straight line. This deviation increases with $t_w$, reaches a maximum and then decreases and finally the PDF again becomes Gaussian. For a given $t_w$ the data are shown both for the positive and negative $\Delta T$. We note that for large $\Delta T$ a small yet definite asymmetry develops in the distribution function~\cite{asymmetry}.

During the denaturation process the observed temperature fluctuation   $\langle(\Delta T)^{2}\rangle/T^{2}$  is much larger than what one would expect from the equilibrium fluctuation. For a microcanonical ensemble at temperature T and heat capacity C the temperature fluctuation is  given by $\frac{\langle(\Delta T)^{2}\rangle}{T^{2}_{eq}} = \frac{k_{B}}{C}$. For the calorimeter at $300K$  the estimated relative fluctuation  $\approx 4.5\times 10^{-22}$. This is several orders less than the observed fluctuation.

We would like to suggest that the observed large fluctuation is a result of coexisting phases that are in dynamical equilibrium. The transition between two phases has energetics associated with it because it involves making and breaking of bonds leading to exchange of energy with the calorimeter. The observed slow dynamics  suggests that the transition between the two phases involves thermal activation across barriers.  The co-existing phases can arise from a number of sources like co-existence of loops (denaturated) and bound segments or due to the system being trapped in mismatched configuration. At this time we cannot definitely identify what leads to these co-existing phases observed in the experiment. We can get an estimate of the average activation energy $\Delta E$ from the wait time ($t_w$) dependence of the fluctuation using $\tau = \tau _0 exp(|\Delta E|/k_BT)$. Here $\tau _0 \simeq 10^{-7}$ s is the microscopic time for a basepair to move together or apart (in the absence of tension or base pairing interactions)~\cite{Cocco}. We find that $\Delta E \simeq 25-30k_BT_0$ where $T_0$ = 300K. The value of $\Delta E$ is similar to the energy at the critical force of unzipping when the DNA is separated using force~\cite{PNAS}. We also note that our observation is distinct from what one expects from co-existing phases in a first order transition. The observation of co-existing phases and the resulting fluctuation is metastable because it dies down after a long wait time. Here the system appears to be trapped for a long time with co-existing phases before it reaches thermal equilibrium. 

To summarize, we have made a direct measurement of temperature fluctuations during thermal denaturation of dsDNA. The fluctuation has a significant low frequency component and it peaks in the temperature region of denaturation. It is absent both in the fully  closed and fully denaturated state. The fluctuation also has a wait time dependence and decays to zero after a sufficient wait time. We suggest that the fluctuation arises due to coexisting phases in dynamical equilibrium.

\section{Acknowledgment}AKR and DC acknowledge financial support from DBT (Government of India) for a sponsored project and Nagapriya acknowledges CSIR for support.


\begin{references}
\bibitem{PB1}D. Poland, H. A. Scheraga, J. Chem. Phys. {\bf 45}, 1456 (1966) and Ibid {\bf 45}, 1464 (1966).
\bibitem{PB2}R. M. Wartell and A. S. Benight, Phys. Rep.  {\bf 126} 67, (1985) 
\bibitem{Bloomfield}C. Richard and A. J. Guttmann, J. Stat. Phys. {\bf 115}, 925 (2004)
\bibitem{Babri}M. Barbi, S. Lepri, M. Peyrard, and N. Theodorakopoulos  Phys. Rev. E {\bf 68}, 061909 (2003)
\bibitem{Garel}T. Garel, C. Monthus and H. Orland, Euro. Phys. Letts. {\bf 55} 132, (2001)
\bibitem{length} Ralf Blossey and Enrico Carlon, Phys. Rev. E, {\bf 68}, 061911 (2003).
\bibitem{Noisetec}Arindam Ghosh, Swastik Kar, Aveek Bid and A. K. Raychaudhuri  {\bf arXiv:cond-mat/0402130} v1 4 Feb 2004. J. H. Scoffield, Rev. Sci. Instr. {\bf 58}, 985 (1987).
\bibitem{water}A.G.W. Leslie, S. Arnott, R. Chandrasekharan and R.L. Ratliff, J. Mol. Biol. {\bf 143}, 49 (1980) 
\bibitem{response}The response of the calorimeter to the bath temperature fluctuations  is that of a low pass thermal filter with roll off at $f^* = \frac {1}{2\pi \tau_{bs}}$ where $\tau_{bs} = R_{th}C_{bath}$ is the time constant of bath to DNA-substrate system. The contribution of the temperature fluctuation of the bath to the  power spectral density (PSD) can be given by 
$S_{Bath}(f) =  \frac{\langle(\Delta T_{bath})^{2}\rangle \tau_{bs}}{1+4\pi^{2}\tau^{2}_{bs}f^{2}}$.
\bibitem{Cocco}S. Cocco, J. F. Marko and R. Monasson, Eur. Phys. J E {\bf 10}, 153 (2003)
\bibitem{PNAS}Claudia Danilowicz, Vincent W. Coljee, Cedric Bouzigues, David K. Lubensky, David R. Nelson and Mara Prentiss, Proc. Natl. Sci. Acad., {\bf 100}, 1694 (2003)
\bibitem{asymmetry}The finite asymmetry in the PDF is a signature of non-equilibrium nature of the system. As the system moves towards equilibrium by absorbing heat from the substrate, there are more cooling jumps than heating jumps.
\end{references}
\end{document}